\documentclass{article}

\usepackage{graphicx} 
\usepackage{booktabs}
\usepackage{cite}

\title{Soft-lockins in Public Sector Acquisitions of Open Source Software-solutions: A Case Study on a Municipal E-Service Platform}

\author{Per Persson \\
  Gothenburg University, Sundsvall municipality \\
  {\underline{ per.persson@ait.gu.se}} \\
  Johan Linåker \\
  RISE Research Institutes of Sweden \\
  {\underline{ johan.linaker@ri.se} }\\}

\begin{document}

\maketitle

\begin{abstract}
\textbf{Background:} Open Source Software (OSS) is often seen as an option to mitigate risks of lock-ins. Yet, single-vendor OSS can still result in soft lock-ins due to knowledge asymmetries and technical barriers. 
\textbf{Aim:} This study explores actors that render such soft lock-ins.
\textbf{Research design:} We conduct a qualitative case study of an E-service Platform (ESP) used by over 190+ municipalities. 
\textbf{Results:} User-driven lock-in factors emerged as a significant category, including limited and non-transparent communication, restrictive qualification requirements in procurement, confusion on maintainership, and comfort in the status quo. Technical lock-in factors include inadequate documentation, dependency management issues, and limited test coverage.
\textbf{Conclusions:} Strong leadership and continuous training is needed to address presence of comfort and conservative culture among municipalities. Open Source Stewards, i.e., neutral hosts for OSS projects, can support municipalities in these tasks while also helping to foster an open, competitive collaboration that can enable a broader supplier ecosystem.
\end{abstract}

\subsubsection*{Keywords:}

Open Source Software, Soft Lock-in, Municipalities, Public procurement, Open Source Steward

\section{Introduction}

Software provides a critical enabler and asset for the digital transformation and provisioning of digital public services by Public Sector Organizations (PSOs) on all levels of government (\cite{mergel2019}). These PSOs typically acquire the software applications and infrastructure components through public procurement and thereby become dependent on the suppliers for successful implementation (\cite{heeks2007}). A significant risk is that this dependency grows so strong that PSOs end up with a lock-in to a specific supplier, technology, platform, or data format (\cite{lundell2020addressing}). Such lock-ins can grow stronger with time and have adverse effects on software quality, innovation, release pace, and cost while also limiting interoperability between systems, access and portability of data, and the ability to make sovereign design decisions in an ever more regionalized global political climate.

Open Source Software (OSS), meaning software available under an OSS license, provides an instrument for pre-empting such lock-ins (\cite{deller2008open}). Ideally, this enables increased competition and market of service provisioning as the source code and knowledge needed to develop and build the source code is open for anyone to consume (\cite{jokonya2015investigating}). 

However, this must not necessarily be the case as the OSS licenses put no requirements on the quality in which an OSS is shared. Boundary resources, such as technical knowledge or specific tools needed to build the OSS, must not necessarily be accompanied. In cases where there is an intention to build a vibrant OSS community and leverage the many benefits of open innovation, this would be standard practice. The case may be different when there is an underlying intention of limiting the possibilities for external contributions and development. The latter is common in single-vendor OSS projects when the vendor (i.e., supplier) aims to retain control while leveraging the openness as a distribution channel and growing a user community (\cite{riehle2012single}). 

In this study, we investigate the phenomena of soft lock-ins to OSS where the user is limited in their sourcing options of a specific single-vendor OSS project due to a knowledge asymmetry between the procurer and OSS supplier. Accordingly, we are interested in understanding what factors may create soft lock-ins, the impact they have, and how they can be avoided. We, accordingly, define our research question as \textit{What factors lead to soft lock-ins to OSS vendors? How can these be mitigated?}

We perform a case study of a Digital E-service Platform (from now on ESP) OSS project developed and maintained as a single-vendor OSS project with a user community of 190+ municipalities within an EU member country. Since the ESP's inception, the municipalities have grown more aware of the soft lock-in that occurs and the implications this has or may have in the future. A qualitative investigation based on interviews, reports, and prolonged engagement from both authors provides a rich narrative presenting several soft lock-in factors that PSOs should be aware of in any acquisition process of OSS. We provide recommendations based on our investigation for how such challenges and risks may be managed and reduced proactively, both from the procurer and OSS supplier perspectives, while enabling a prosperous collaboration and a sustainable OSS project.


\section{Background and related work}

\subsection{Public procurement and development of OSS}

The adoption of OSS and procurement of OSS-based solutions and related services is thoroughly reported in the literature, especially at the municipal level of government. A main driver for the adoption is associated with potential cost savings (\cite{ven2012qualitative, allen2010open, tosi2015surveying, koloniaris2018possibilities, deller2008open}) while also lower switching costs (\cite{ven2012qualitative}), and potential to avoid lock-in are highlighted (\cite{deller2008open}).

Against the backdrop, several challenges and barriers commonly impede OSS adoption. From a market perspective, the lack of mature and high-quality alternatives compared to proprietary and existing options limits the choice for the PSOs (\cite{magnusson2011intentions, petrov2018barriers, gurusamy2011case, tosi2015surveying}). 

Internally, the PSOs, resistance to change, risk aversiveness, and preference to preserve the status quo with existing solutions reported cultural barriers to adoption (\cite{bouras2013methodology, petrov2018barriers}). There is typically a need for strong managerial support to enable adoption and to effectively guide departments through the cultural and technical barriers (\cite{petrov2018barriers, hamid2016framework, tosi2015surveying, silic2017open, van2015adopting, shaikh2016negotiating}). 

Limited internal technical capacity or experience of OSS adoption is commonly reported, highlighting the need for internal training (\cite{bouras2013methodology, gurusamy2011case, hamid2016framework, koloniaris2018possibilities}) and champions that drive organizational change (\cite{shaikh2016negotiating}), but also provide a bridge towards the external OSS communities and the larger ecosystem of actors (\cite{van2015adopting}). The latter can be compared to the emerging phenomena of public sector Open Source Program Offices (OSPOs), i.e., centers of competency that can support OSS adoption and best practice within or across organizations.

The availability of commercial support for an OSS project is highlighted in several studies as a key requirement for adopting any OSS to ensure quality, security, and use~\cite{bouras2013methodology, holck2005managerial, ven2009importance}. Yet, the availability of suppliers is also commonly considered a challenge due to limited availability (\cite{gurusamy2011case, magnusson2011intentions}). An alternative is to develop the necessary technical capabilities internally (\cite{holck2005managerial}), although this is complicated, especially due to limited budgeting and a general reliance on outsourcing.

While commonly adopted in the industry and wider OSS ecosystem, an emerging practice is that PSOs create OSS stewards, i.e., dedicated neutral organizations that host the projects and facilitate their open, collaborative development of them. Viseur and Jullien describe the case of the CommunePlone OSS project and how Wallonian municipalities collaborate through a co-owned service supplier (\cite{viseur2023communesplone}). Perhaps most mature, Frey presents how the Danish municipalities collaborate on the requirements definition, procurement, and governance of joint OSS projects through OS2, an independent association with 80+ out of 98 municipalities in Denmark as members (\cite{frey2023we}).

In terms of lock-in, literature primarily reports on cases where PSOs state references to specific and proprietary software applications, data formats, or other forms of non-open IT standards or trademarks that effectively limit the number of products, services, and providers that can be procured (\cite{lundell2021enabling, lundell2016standarder}). Failing to collect necessary third-party licenses, as well as defining exit strategies, are other common issues (\cite{lundell2020addressing}). Active engagement and contributions to OSS projects are highlighted as a means of breaking supplier dependence and pre-empting any lock-in effects (\cite{blind2019relationship}).

\section{Research design}

The study uses a design science research approach where the intent is to generate design knowledge that can be tailored and applied as a solution by practitioners in the real-world problem context (\cite{venable2006}). The approach entails an iterative transition from studying the problem context towards designing a solution proposal, which is then applied and possible to evaluate in the problem context. 

A case study design is chosen as the phenomena are deeply entangled in the problem context and difficult to isolate and control (\cite{runeson2012}). The case was sampled by convenience due to both authors' in-depth knowledge and, to various degrees, prolonged engagement (\cite{easterbrook2008}) with the case. The openness of the ESP OSS project is chosen as our unit of analysis, where we focus specifically on identifying the factors that in any way create a knowledge asymmetry to the benefit of the vendor and, by extension, a soft lock-in for the users of the OSS, limiting their ability in developing, building, and using the OSS, either through internal or procured resources.

An initial theoretical framework of Lock-in factors, their implications, and potential mitigation tactics was developed based on reports and documentation (e.g., meeting minutes from the municipal user association). A questionnaire was then designed based on the framework to validate, reject, extend, and enrich it through semi-structured interviews with individuals of different perspectives and relation to the ESP OSS project.

We, accordingly, performed a total of nine interviews with eight interviewees selected through purposeful sampling (\cite{patton2014}) to provide complementary and overlapping perspectives of the case. Demographics of the interviewees and rationale for being selected are provided in Table\ref{tab:interviewees}. Each interview was conducted with both authors present through online video platforms, lasting about 60 minutes each. The interviews were recorded and transcribed using automated tool-support and later manually processed for quality and consistency. A summary of each interview and corresponding transcript was communicated to the respective interviewees for member-checking.

Each transcript was then coded by the first author using open coding (\cite{saldana2021}) together with our initial theoretical framework as an a-priori code book. The list of codes was iteratively revised and extended as each interview was analyzed, as well as codes describing the evolution of the ESP OSS projects and significant events occurring during the period. 

The second author also coded the first two interviews, after which there was a joint discussion addressing any disagreements. For the remaining interviews, the first author performed the main coding while continuously discussing these with the second author through peer-debriefing (\cite{easterbrook2008}). Codes were further continuously aggregated into higher-level categories of axial coding (\cite{saldana2021}). A final version of the theoretical framework was communicated to all interviewees for a last round of member checking and feedback. There was a general agreement on the framework's content, and only minor adjustments were incorporated. The final version is presented in the Findings section.

\begin{table*}[tbh!]
\caption{Overview of interviewees, their demographics, and the rationale for inclusion.}
\label{tab:interviewees}
\begin{tabular}{p{0.7cm}p{2.5cm}p{3cm}p{4.2cm}}
\hline
\textbf{ID} & \textbf{Role} & \textbf{Experience} & \textbf{Connection to the ESP OSS}  \\ \hline

I1 &  
IT strategist, Municipality&
25+ years in  municipalities IT and digitalization&  
Participant in the EU project initiating the ESP OSS. Driving force in its continued development.\\ \midrule

I2 &  
 ESP user, Municipality&
 Six years experience working with the ESP. Chairman of the ESP user group since 2018&  
Customer to the main supplier of the ESP SaaS offering, chairman of the ESP user group, and engaged in the future evolution of the platform.\\ \midrule

I3 &  
 Developer, consultant&
 20+ years of experience in system development&  
Developer that raised early issues with contributing to the platform.\\ \midrule

I4 &  
 ESP user, Municipality&
 User and Board member of user association.&  
Customer to the main supplier of the ESP SaaS offering and board member in the ESP user group.\\ \midrule

I5 &  
 Developer, Municipality&
 20+ years experience in system development&  
System developer and main contributor to the 2022 report.\\ \midrule

I6 &  
 IT architect, Municipality&
 20+ years in  municipalities IT development &  
Customer to the main supplier on the ESP SaaS offering. Engaged in the evolution of the platform.\\ \midrule

I7 &  
 CIO, IT service supplier&
 25+ years in the software development business&  
CIO of the company that compiled the 2019 report.\\ \midrule

I8&  
Product manager, Main supplier&
 15+ years in the software development business&  
Founder of the main supplier company and creator of the ESP.\\ \midrule

\end{tabular}
\end{table*}

\section{Threats to validity and Limitations}
Internal validity threats include bias in data collection, as participants' responses may be influenced by their personal experiences and positions. Researcher bias could also be present due to the authors' prolonged engagement with the case. To mitigate these biases, we used member-checking and peer-debriefing techniques, but some degree of bias may remain.

External validity concerns the generalizability of our findings, which are based on a specific case of an ESP within the municipal sector within a European country. Therefore, the results may not be directly applicable to other contexts, regions, or types of OSS projects. Additionally, the purposeful sampling of interviewees, while providing rich insights, may not represent the full spectrum of perspectives within the user community or the broader public sector.

The scope of our study is limited to a single OSS project within a specific sector. While the depth of our case study provides valuable insights, it limits the breadth of coverage. Future research could expand to include multiple OSS projects across various sectors and regions to enhance the comprehensiveness of the findings. 

Cultural and organizational dynamics specific to the concerned country's municipalities might influence the findings. Comparative studies involving different cultural and organizational contexts could provide a more nuanced understanding of the factors contributing to soft lock-ins in OSS projects. By addressing these threats to validity and acknowledging the limitations, we aim to provide a balanced and transparent account of our research. These considerations highlight the robustness of our findings and pave the way for future research to build upon and extend our work, ultimately contributing to a deeper understanding of soft lock-ins in OSS acquisitions within the public sector.

\section{Case Timeline and Narrative}

Below, we present a descriptive summary of the inception and evolution of the ESP OSS project. 

\subsection{Project Inception}

The E-Service Platform \footnote{An E-Service Platform provides digital case registration for external (citizens and entrepreneurs) or internal (employees) users. Creating forms for data-collection for cases is configurable, and the cases are stored in a database. Case managers can then pick up cases and, most often, insert the cases into department-specific case management support systems for further management} (ESP) was established in a joint project with five municipalities in 2012 financed by the European Union. The goal of the project was to make it easier for companies and citizens to gain access to information and communicate with the municipality on issues that mainly concerned urban planning and building permit issues. 

The ESP, accordingly, implemented twelve e-services in support parts of these tasks. Two suppliers were used in the development process. One supplier developed a smaller map-based plugin, while the other (from now on, the \textit{main supplier}) developed the overarching platform. The ESP went into production in 2014.



ESP was published under the AGPL 3.0 OSS license, with the ambition that suppliers who use the code base and offer the ESP to customers as a cloud service must contribute back to the OSS project with the further development they are doing in the platform (I1 and I8). When the EU-funded project closed, one of the participating municipalities was given the copyright ownership of the developed code and a foundation that was already maintaining the OSS Java Framework the ESP was based on and relied on maintaining the ESP as well (I1).

Following that, the main supplier who built the platform in the project created an offering hosting ESP as a SaaS, offering operations and management at a fixed price and further development via traditional procurement (I1 and I8). The supplier is the only one that delivers ESP as a SaaS for the reasons described below. ESP is today the most popular e-service platform amongst the concerned country's municipalities and is currently used by approximately 65 percent of the concerned country's municipalities that work together through an informal user association (I1, I2, and I4).

\subsection{Project Evolution}
After the launch of the platform, voices started to be raised from other suppliers, apart from the main supplier, about issues that presented obstacles to contributing to the platform development (I3). In parallel, platform users started to raise issues about the main supplier's delivery capacity and the lowering development pace (I3, I4, and I6). 

The issues escalated over the years and were discussed in a yearly meeting of the user association in 2018, during which the topic “future-proof management of ESP” was selected as the most important topic to discuss. One user association member expressed that \textit{``with the large number of municipalities that now use the platform, problems with, among other things, delivery times are being noticed. That leads to increased risks for the municipalities. It was raised that the feeling is that small municipalities are sometimes given lower priority by the supplier than large municipalities due to lack of resources''} (Meeting memos, 2018).

The conclusion was that documentation and access to code are key to a functioning openness and transparency, which can ultimately create an open market. Several proposals for measures were presented, including the creation and publication of a software architecture description and rules and routines for documentation of code. 

It was proposed that all municipalities should require that applicable rules and procedures for the documentation of code must be followed when procuring the development of ESP. Also, comprehensive documentation on how a development environment is set up should be created and published. All participants at the meeting stood behind these measures and it was decided to continue the work on this in a committee.

In 2019, a new possible supplier showed interest in contributing to the platform yet found it difficult (I7). Challenges were summarized in a report and presented to the leading municipality regarding the development of the platform and to the ESP user association. The main concerns presented in the report regarded limitations in documentation (e.g., in setting up the system and development environment), a lack of dependency management, no automated tests, and code quality not being analyzed. Documentation was addressed by the main supplier following the report (I2 and I5).

In 2022, as a mitigation to the escalating long delivery times from the main supplier, a municipality with an in-house development department aimed to contribute to the platform. They arrived at similar conclusions as the aforementioned report from 2019, leading to the user association initiating a more focused discussion, both internally and with the main supplier, on how the concerns raised best can be addressed (I2 and I6).

\section{Lock-in factors}
In our studies, we found two main types of lock-in factors: user-driven and technical. A presentation of each category with suggested mitigation tactics follows below, along with a summary in table\ref{tab:lockinfactors}.

\begin{table*}[tbh!]
\caption{Overview and brief descriptions of the user-driven and technical lock-in factors observed in the study.}
\label{tab:lockinfactors}
\begin{tabular}{p{2cm}p{9.3cm}}
\toprule
\multicolumn{2}{l}{\textbf{User-driven lock-in factors}} \\ \midrule
Communication & Limited communication between municipalities and towards the main supplier. Low level of transparency and awareness of developed functionality and ongoing or planned work. \\
Procurement & Inconsistent and disqualifying qualification requirements on suppliers limiting the number of potential bidders to incumbents. \\
Maintainership & Mixed opinions on who is responsible for the maintenance of the ESP OSS project, and for facilitating any open collaborative development and community management. \\
Comfort & A comfort and risk-aversiveness in preserving status quo. Preference to risks implied by technical debt and soft lock-in, before what could come in an unknown future. \\ \midrule
\multicolumn{2}{l}{\textbf{Technical lock-in factors}} \\ \midrule
Dependency management & A limited overview of what dependencies that exist towards third-party components, or what version of dependencies have been included.\\
Development environment & Lock-in to a specific Integrated Development Environment due to technical design choices, e.g., dependency management. \\
Documentation & Limited documentation regarding the development, contribution and onboarding, build-environment, and running the ESP.\\
Testing & No tests are present in the OSS project code base, e.g., unit tests, functional tests, and end-to-end tests.\\
Code quality & Lack of visualized automatic analysis of the source code, searching for errors, security issues, and estimating code quality\\ \bottomrule
\end{tabular}
\end{table*}

\subsection{User-driven lock-in factors}

User-driven lock-in factors consider practices, organization, and culture among the users (i.e., municipalities), either in isolation or collectively in the user community that triggers a self-imposed lock-in effect towards the supplier.


\subsubsection{Communication} 

According to the interviewees, there is a lack of transparency and awareness of what is requested or procured from the different municipalities towards the main supplier (I3, I5-7). Typically, the larger and more resourceful municipalities are reported to make requests directly to the main supplier, which is not recorded or communicated in any way to the rest of the user community (I2, I6, and I8). The main supplier takes note and tries to gather municipalities that have made similar requests. 

Still, several examples are brought up of how municipalities are made aware of functionality that has been procured and implemented that they are both interested in and would have liked to participate both in the requirement definition phase and co-funding. By effect, it becomes difficult to maintain a common vision and road-map for the ESP, implementations risk becoming user-specific rather than general, and the cost is taken on by single municipalities, which could otherwise have been shared more collectively (I2-4, I6). The main supplier also requests a more organized and transparent collaboration with the municipalities as this would decrease overhead and increase development efficiency, further benefiting the customers.


\textbf{Mitigation tactic:} Several interviewees suggest the user association could strive to take on a more formal responsibility in promoting and facilitating open and systematic communication within the community and with the main supplier. By sharing and discussing potential requirements, e.g., in a common issue tracker, all users are enabled to consider whether they are interested in partaking in an acquisition process, and are made aware of the potential functionality. This would also create more effective and joint communication with the main supplier, thereby avoiding the risk of isolated and side-track discussions.

\subsubsection{Procurement} 

Concerns were expressed that procurement practices effectively helped limit the onboarding of additional suppliers. One of the potential suppliers that attempted to enter the market for the ESP (I7) specifically raised the use of qualification requirements as a barrier. There was typically a minimum requirement of having between two to twenty earlier customer cases, which proved impossible to meet as the existing main supplier was the only one that could meet such demands. Further, when an agreement is about to reach its end date, a user typically tends to prolong the existing contract instead of creating a new tender and opening up for competing bids.

\textbf{Mitigation tactic:} As raised by the interviewees, procurement procedures need to harmonize and specifically take note of how new suppliers can be enabled to compete for new tenders, including those with limited or no previous experience.

\subsubsection{Maintainership} 

The municipalities that took part in the initial stages of the ESP's evolution had limited knowledge and capacity to take on the technical maintainership of the code base, e.g., peer-reviewing and accepting code contributions from the suppliers. These tasks were instead outsourced to the main supplier, who, on request, would do code reviews of other suppliers' contributions (I1 and I8). From the main supplier's perspective, the municipalities still had responsibility for the maintainership of the ESP (I8), while the view differs among municipalities and competing suppliers. Regardless, I1 and I7 describe discomfort and unwillingness among the competing suppliers to have their code contributions reviewed by a competitor, which, by effect, further concentrated development efforts on the main supplier. The main supplier notes that they have thus far not received any requests for code reviews of other suppliers' contributions. 

\textbf{Mitigation tactic:} Interviewees stressed the need for the user association to take an active and more formal role in the maintenance and governance of the ESP, ideally through a common and trusted entity. References were made to the Danish municipal association OS2, which acts as a steward for the OSS projects initiated and collaboratively developed by its members. The user association could accordingly search for or establish a similar and more formal organization within the national context or look to more general and international OSS foundations such as the Eclipse Foundation.

\subsubsection{Comfort} 

The users (i.e., municipalities) are pleased overall with the ESP, as well as with the main supplier, who is described as competent and service-minded. The ESP is judged as more stable and secure than many other systems used in the municipality sector. This aspect outweighs the decreased development pace, which has been noted during past years and by some described as increasing (I2, I4, and I6). This trade-off can be explained by a comfort factor widely spread among the users - having a good dialogue in the user association and with the main supplier is considered more valuable than increasing the development pace of the ESP. By extension, there is limited interest in interrupting the status quo and in attempting to onboard new suppliers as there is a fear that such a change might worsen the current relationship with the main supplier and perceived comfort from the situation.

\textbf{Mitigation tactic:} Training on how to consider OSS in an acquisition process and collaborate on its development is highlighted by several interviewees. Benefits, as well as risks and potential impacts, need explanation, and how they can be balanced and addressed. Further, there needs to grow an open culture within and across the municipalities through the user association that fosters an open collaboration and grows the courage to question the comfort of the situation.


\subsection{Technical lock-in factors}

Technical lock-in factors consider technical issues acting as impediments for a developer not earlier familiar with the ESP to be able to set up a development environment and contribute to the project. A synthesis of the reports from 2019 and 2022, independently summarizing these technical lock-in factors, confirmed by I3, and I5-7 is presented below. This is contrasted with the view of the main supplier (I8) accordingly.

\subsubsection{Dependency management} 

A limited overview has been observed of what dependencies exist from the ESP towards external packages or what version of dependencies have been included. A reported consequence of the limitations in dependency management has been that users have felt prevented from or challenged in building and updating a version of the system that can be trusted, either in its current or future state. While dependency management is currently managed through the Eclipse IDE, there have been expressed wishes that these should rather be managed through an external package manager such as Maven (I3, I6-7). The main supplier (I8), in response, considers Eclipse to be a convenient system for managing dependencies, and there has been no formal request by its customers to change the system (e.g., to Maven). I8 further explains that they revised system documentation upon request in 2019, including how dependencies are managed, yet received no feedback until 2022, when a new report concluded the same issues. These limitations in communication between municipalities and the main supplier (as confirmed by I1) further emphasize the need for a more formal and organized maintainership and collaboration within the user association.

\textbf{Mitigation tactic:} Several interviewees and reports converge on the need to require all dependencies to be declared in an OSS project, preferably using standard package managers (e.g., Maven), and towards components that are publicly available under an OSS license, and that can be built using OSS tools. Such wishes and requirements need to be explicitly defined in contracts between the customers and suppliers.

\subsubsection{Development environment} 

The Eclipse Integrated Development Environment (IDE) has been used throughout the development of the ESP by the main supplier (I8). The reports and other interviewees have raised concerns regarding the lock-in nature of the IDE, e.g., due to the dependency management that is integrated into the Eclipse IDE. Concerns highlighted include barriers to onboarding new developers with limited knowledge of the current IDE (Eclipse IDE), general preferences for alternatives otherwise available, and concerns also understood by the main supplier. The main supplier considers the choice of IDE as their internal preference and highlights that it is an internationally recognized system with an active developer and user community. Documentation is reported to have been improved in their 2019 revision of the documentation (I8), although with concerns reiterated in the 2022 report.

\textbf{Mitigation tactic:} There should be a consistent and explicit requirement from the customer side if a specific development environment should be used or be compatible with. Further, documentation should be requested and validated to ensure the required level of usability and setup of the development environment.

\subsubsection{Documentation} 

Limitations in documentation are reported as an additional factor impacting the possibility for new developers to on-board the project, e.g., in terms of architectural design, instructions on how to set up and compile the project, the development process, and guidelines on how to contribute to the OSS project. The experience is expressed both by developers inside of the municipalities and competing suppliers attempting to set up a development and hosting environment. 

The main supplier understands the need for the documentation but limits its responsibility to provide documentation on request, referring to the fact that they are not the maintainers of the OSS project. They further note that documentation was improved on request in 2019 but that no response was received until 2022, when a new report was performed, concluding with similar concerns.

\textbf{Mitigation tactic:} As a minimum, a System Analysis and Design (SAD) document should be required, along with development and operations instructions and a continuously updated change log. It is further desirable to provide documentation of best practices, guidelines, and rules that apply to development.

\subsubsection{Testing} 

A general lack of test cases in the OSS project code base has been reported (e.g., unit tests, functional tests, and end-to-end tests). Interviewees (I3, I6-7) consider the lack of test cases as a risk in terms of quality and security and as a barrier to contributions as there they would be afraid of introducing bugs and errors without knowing. Interviewees associate the limited test coverage with an elevated cost and risk since manual tests are performed late in the project life cycle, take a long time to perform, and there is a risk that they are performed incorrectly (or not performed at all) due to the human factor. The main supplier (I8), in turn, explains that tests on all levels are performed internally before releases, and that the limited test coverage in the code base is due to the modular structure of the ESP, and that the number of configurations is beyond what would be feasible to test for. They highlight that no explicit requirements have been made concerning test coverage and further note that their customers (i.e., the municipalities) are covered by the contracts and service level agreements where the supplier is responsible for the quality of the ESP where provided as a service.

\textbf{Mitigation tactic:} Interviewees suggest that automated tests should be introduced on all levels (e.g., unit tests, integration tests, etc.) with a goal of test coverage as near 100 percent as possible, which can detect problems earlier in the project life cycle. If the combinatorial complexity raised by the main supplier provides a barrier, automated tests may still be developed for a selected set of representative configurations. To fully support the testability of the solution, there is a need to overhaul the system design and ensure the solution is properly modularized (\cite{baldwin2000design}) and designed for extensibility and backward compatibility (\cite{morgan2000cathedral}).

\subsubsection{Code quality} 

Report and interviewees (I3, I6-7) have explicated concerns regarding the lack of visualized automatic analysis of the source code, searching for errors, security issues, and estimating code quality. By effect, they note that it becomes hard to trust the software and presents obstacles to contribute. References are drawn to how social development platforms such as GitHub provide a corresponding interface, also allowing for scanning functionality to be automated as part of a build process. The main supplier responds that they are performing these analyses in-house but do not expose the results. The source code is currently managed on an SVN repository owned and managed by one of the leading municipalities in the user association. Discussions have, however, been initiated to mirror and, with time, move the source code management to GitHub.

\textbf{Mitigation tactic:} Automated source code analysis, e.g., in terms of security and quality, should be requested and defined in the contract between customers and suppliers. Similar requirements regarding the use of development platforms (e.g., GitHub) and the use of any specific scanning functionality should, in similar ways, be explicitly requested.

\section{Discussion and Conclusions}

\textbf{User-driven Lock-in:} User-driven factors are created and reinforced by the users themselves and originate from the municipalities' means of collaborating and communicating throughout the project's evolution. The differing views on who owned the formal maintainership of the project, along with the original lack of capacity to take on the role, played a significant part in leading up to the soft lock-in. Everyone was pointing at each other. The contract with the main supplier, enabling them to do code reviews on request, created an unwillingness for potential competing suppliers to engage. Procurement practice with strict qualification requirements effectively reinforced the lock-in to the level otherwise found in more traditional lock-in situations, e.g., to commercial standards or technologies (\cite{lundell2016standarder}).

\textbf{Technical Lock-in:} The soft lock-in was further reinforced by the technical barriers, which we conjecture to a large degree are related to and enforced by the limitation in collaboration and communication between the municipalities and with the main supplier. Interviewees are in rather agreement that the technical barriers such as dependency management, inadequate documentation, development environment setup issues, and lack of testing practices impede new developers from contributing to the ESP. Adding new developers could facilitate a higher development pace of the ESP, even though the main supplier highlights that with more suppliers adding to the same code base, the need for internal administration would increase, putting efforts into verifying other suppliers' contributions before accepting them to their delivery. A way to mitigate that would be to adopt best practices regarding OSS development (\cite{fogel2005producing}) and community management (\cite{bacon2012art}) and move towards a more open collaborative development model. 


\textbf{Open Source Stewards:} One solution, as raised by interviewees both from the municipalities and the mains supplier, regards the establishment of an independent entity or leveraging an existing OSS foundation to oversee maintainership. Such a transition could help to improve and facilitate an open collaboration both between the municipalities and toward suppliers and ensure, e.g., the presence of the necessary documentation, tests, and onboarding processes. The steward would act as a joint champion for the municipalities and serve as an interface to the wider ecosystem of actors concerning the OSS project (\cite{van2015adopting, shaikh2016negotiating}), and ensuring that both the municipalities' needs and the main supplier and prospective new suppliers' business models are considered.

Similar approaches have proven successful on the municipal level both in Denmark (\cite{frey2023we}) and Belgium (\cite{viseur2023communesplone}). Technical resources can be pooled, road mapping planned collectively, and features planned, co-funded, and collectively procured. The more inclusive and competitive environment would enable further suppliers to join the community and provide services complementary to the main supplier, an idea the main supplier (I8) proved open to.

\textbf{Comfort and Conservative Culture:} Preempting and addressing the user-driven lock-in factors, however, goes beyond the creation of a joint steward for the OSS project. The case study highlights the impact of culture regarding comfort, risk-aversiveness, and preference for preserving a status quo, commonly reported issues in the municipal context (\cite{bouras2013methodology, petrov2018barriers}). The uncertainties of what could come out if the balance is changed are perceived as scarier than what the risks of the current situation may imply in the future. 

A general recommendation from our study is, therefore, to proceed slowly with care but with determination. Strong leadership is needed when addressing cultural issues at hand (\cite{hamid2016framework, tosi2015surveying}), and effort is invested into information sharing and education about the situation that needs to be solved (\cite{koloniaris2018possibilities}). Collaborating closely with present suppliers should be preferred and promoted, as these are needed for the creation of a sustainable OSS community. In the case of the ESP, the main supplier is the main reason for the success of the ESP in terms of adoption and collaboration among the concerned country's municipalities, which accounts for 65 percent of the total. Interviews further point to the fact that the supplier is very much open to a more open collaborative development model and community, but that it is the municipalities that need to take on the role of maintainership.

\textbf{Future work:} There is a need for future investigations into how municipal and public sector collaboration on OSS can be enabled and supported within the confines of an OSS steward, public sector specific such as OS2 or general such as the Eclipse or Linux Foundations. Research should further consider how such stewards can enable cross-border collaboration of OSS projects, although many OSS projects may be expected to be country-specific, e.g., due to national regulations, processes, and culture. Additionally, further research is needed specifically into how open collaborative development can be bridged with public procurement practice and regulation, as many PSOs (especially on the municipal level) do not have the necessary technical capacity internally.

\bibliographystyle{splncs04}
\bibliography{main}

\end{document}